\documentclass[twoside]{article}
\usepackage{amssymb}
\def\R{{\mathbb R}}
\def\E{{\mathbb E}}
\def\N{{\mathbb N}}

\def\C{{\mathbb C}}

\newtheorem{theorem}{Theorem}

\title{Nontrivial models of quantum fields with indefinite metric}
\author{Sergio Albeverio and Hanno Gottschalk\\ Institut f\"ur angewandte Mathematik\\
Rheinische Friedrich-Wilhelms-Universit\"at Bonn\\
D-53115 Bonn, Germany\\ albeverio@uni-bonn.de/gottscha@wiener.iam.uni-bonn.de}

\begin{document}
\maketitle
\pagestyle{myheadings}
\thispagestyle{empty}
\markboth{S. Albeverio, H. Gottschalk}{Nontrivial models with indefinite metric}


\section*{Introduction}
The non perturbative construction of quantum field models with nontrivial scattering in arbitrary dimension $d$ of the underlying Minkowski space-time
is much more simple in the framework of quantum field theory with indefinite metric than 
in the positive metric case. In particular, there exist a number of solutions in the physical dimension
$d=4$, where up to now no positive metric solutions are known. Here we review, why this is so, and we discuss some examples 
obtained by analytic continuation from the solutions of Euclidean covariant stochastic partial differential equations (SPDEs) driven by non-Gaussian white noise.

\section*{Main text}
It has been proven by F. Strocchi that a quantum gauge field in a local, covariant gauge cannot 
act on a Hilbert space with a positive definite inner product. But it is possible to overcome this obstacle by
passing from a Hilbert space representation of the algebra of the quantum field to Krein space representations 
in order to preserve locality and covariance under the Poincar\'e group. 

A Krein ${\cal K}$ space is an inner product space which also is 
a Hilbert space with respect to some auxilary scalar product. The relation between the inner product $\langle.,.\rangle$ and the auxilary scalar product
$(.,.)$ is given by a self-adjoint linear operator $J: {\cal K}\to{\cal K}$ with $J^2={\bf 1}_{\cal K}$ and $\langle.,.\rangle=(.,J.)$. $J$ is called the 
metric operator. A quantum field acting on such a space is called a quantum field with indefinite metric. The formal definition is as follows:

Let ${\cal D}\subseteq {\cal K}$ be a dense linear space and $\Omega\in{\cal D}$ a distinguished vector henceforth called the vacuum. Let ${\cal S}={\cal S}(\R^d,\C^N)$ be the space of Schwartz
test functions with values in $\C^N$. A quantum field $\phi$ by definition is a linear mapping from ${\cal S}$ to the linear operators on ${\cal D}$. One usually assumes that ${\cal D}$ is generated as the linear span of vectors generated by repeated application of field operators to the vacuum.
The following properties should hold for the quantum field $\phi$: 

\begin{enumerate}
\item (Temperednes) $f_n\to f$ in ${\cal S}$ $\Rightarrow $ $\langle\Psi,\phi(f_n)\Phi\rangle\to\langle\Psi,\phi(f)\Psi\rangle$ $\forall \Psi,\Phi\in {\cal S}$;
\item (Covariance) $\exists$ a weakly continuous representation $U$ of the covering of the orthochronous, proper Poincar\'e group $\tilde P^\uparrow_+$ by linear operators on ${\cal D}$ which is $J$-unitary, i.e. $U^{[*]}=U^{-1}$ with $U^{[*]}=J U^*J|_{\cal D}$ and leaves $\Omega$ invariant. $\phi$ is said to be covariant with respect to $U$ and a representation
$\tau$ of the covering of the orthochonous, proper Lorentz group $ \tilde L^\uparrow_+$ if $U(g)\phi(f)U(g)^{-1}=\phi(f_g)$, where $f_g(x)=\tau(\Lambda)f(\Lambda^{-1}(x-a))$, $g=\{\Lambda,a\}$, $\Lambda\in \tilde L_+^\uparrow$, $a\in\R^d$; 
\item (Spectrality) Let $U(a)$, $a\in\R^d$, be the representation of the translation group and let $\sigma=\cup_{\Psi,\Phi\in{\cal D}}\mbox{\rm supp}{\cal F}(\langle\Psi,U(.)\Phi\rangle)$ with $\cal F$ the Fourier transform (in the sense of tempered distributions). Formally, $\sigma$ is the joint spectrum
of the generators of space-time translations $U(a)$. The spectral condition then demands that $\sigma\subseteq \bar V_0^+$ the closed forward lightcone in energy-momentum space;
\item (Locality) There is a decomposition $\C^N=\oplus_\kappa V_\kappa$ such that for each $f,h\in{\cal S}$ taking values in one of the $V_\kappa$ and having space-like separated supports we either have $[\phi(f),\phi(h)]=0$ or $\{\phi(f),\phi(h)\}=0$, where $[.,.]$ is the commutator and $\{.,.\}$ the anti-commutator;
\item (Hermiticity) There is an involution $*$ on ${\cal S}$ such that $\phi(f)^{[*]}=\phi(f^*)$.  
\end{enumerate}

The quantum mechanical interpretation of the inner product of two vectors in ${\cal K}$ as a probability amplitude however gets lost. It has to be restored by
the construction of a physical subspace of ${\cal K}$ where the restriction of the inner product is non-negative. This is called the
Gupter-Bleuler gauge procedure. Typically, one considers the problem of constructing
quantum fields with indefinite metric first. This is being seen as the dynamic side of the problem. The construction of the physical states,
which can also be seen as implementation of quantum constraints, is often postponed.

The vacuum expectations values (VEVs), also called Wightman functions, of the quantum field theory with indefinite metric (IMQFT) are defined as:
\begin{equation}
\label{1eqa}
W_n(f_1\otimes\cdots\otimes f_n)=\langle\Omega,\phi(f_1)\cdots\phi(f_n)\Omega\rangle\, ,~~f_1,\ldots,f_n\in{\cal S}.
\end{equation}

An axiomatic framework for (unconstrained) IMQFT has been suggested by G. Morchio and F. Strocchi in terms of the Wightman functions $W_n\in {\cal S}'$, $n\in\N_0$.
 Previous work on the topic had been done by J. Yngvason. 
These generalized Wightman axioms of Morchio and Strocchi replace the positivity condition on the Wightman functions by a so-called Hilbert space structure
condition  (HSSC):
For $n\in \N_0$ $\exists p_n$ a Hilbert seminorm on ${\cal S}^{\otimes n}$  such that
\begin{equation}
\label{2eqa}
|W_{n+m}(f\otimes h)|\leq p_n(f)p_m(h)~~\forall n,m\in \N_0, ~f\in{\cal S}^{\otimes n},~h\in{\cal S}^{\otimes m}.
\end{equation}
 This condition makes sure that a field algebra on a Krein space with VEVs equal to the given set of Wightman functions can be constructed. The remaining axioms of the Wightman framework --
temperedness, covariance, spectral condition, locality and Hermiticity -- remain the same. Clustering of Wightman functions is assumed at least for massive theories:
\begin{equation}
\label{3eqa}
\lim_{t\to\infty}W_{n+m}(f\otimes h_{ta})=W_n(f)W_m(h)~~\forall n,m\in \N_0, ~f\in{\cal S}^{\otimes n},~h\in{\cal S}^{\otimes m}, 
\end{equation}
for space-like $a\in\R^d$. It fails 
to hold in certain physical contexts where multiple vacua (also called $\Theta$-vacua) accompanied with massless Goldstone Bosons occur due to spontaneous symmetry breaking. 

In the original Wightman axioms there are essentially two nonlinear axioms: Positivity and clustering. Non-linear here means that checking that condition involves more than one 
VEV with a given number of field operators. The cluster condition can be linearized by an operation on the Wightman functions called 'truncation'. The equations
\begin{equation}
\label{4eqa}
W_n(f_1\otimes\cdots\otimes f_n)=\sum_{I\in{\cal P}^{(n)}}\prod_{\{j_1,\ldots.j_l\}\in I\atop j_1<j_2<\ldots<j_l}W_n^T(f_{j_1}\otimes\cdots\otimes f_{j_l})
\end{equation}
recursively define the truncated Wightman functions $W_n^T$ for $n\in\N$. Here ${\cal P}^{(n)}$ stands for the set of all partitions of $\{1,\ldots,n\}$ into disjoint, nonempty sets. 
 Unfortunately, the positivity condition
(at least when combined with nontrivial scattering) becomes highly non-linear for truncated Wightman functions. This can be seen as one explanation, why it is so difficult to find non trivial (i.e. corresponding to nontrivial interactions) 
solutions to the Wightman axioms. 

But it turns out that in contrast to positivity the Hilbert space structure condition is essentially linear for truncated Wightman functions.
\begin{theorem}
\label{1theo}
If $\exists $ a Schwartz norm $\|.\|$ on ${\cal S}$ such that $W_n^T$ is continuous with respect to $\|.\|^{\otimes n}$ for $n\in\N$ then the associated sequence of Wightman functions $\{W_n\}$ fulfills
the Hilbert space structure condition (\ref{2eqa}).
\end{theorem}
Note that $\|.\|^{\otimes n}$ is well-defined as ${\cal S}$ is a nuclear space.
 This theorem makes it much more easy to 
construct quantum fields with indefinite metric. In particular, all known solutions of the linear program for truncated Wightman functions lead to an abundance
of mathematical solutions to the axioms of IMQFT, as long as the singularities of truncated Wightman functions in position and energy-momentum space do not become stronger and stronger with growing $n$. E.g. the perturbative solutions to Wightman functions of A. Ostendorf and O. Steinman provide solutions when
the perturbation series is truncated at a given order.

In the classical work on constructive quantum field theory relativistic fields in space-time dimensions $d=2$ and $3$ have been constructed by analytic continuation from Euclidean random fields. This in particular has
led to firm connections between quantum field theory and equilibrium statistical mechanics. 
Let us discuss one specific class of solutions of the axioms of IMQFT for $d$ arbitrary which also stem from random
 fields related to an ensemble of statistical mechanics of classical, continuous particles. On the mathematical side, this is conneted with using random fields with Poisson distribution. As usually in constructive QFT, the moments, also called Schwinger functions, of the random
field can be analytically continued from Euclidean imaginary time to relativistic real time. That this is possible results from an explicit calculation. Axiomatic results cannot be used as they depend on positivity or reflection positivity in the Euclidean space-time, respectively. 
 
By definition, a mixing Euclidean covariant random field $\varphi$ is an almost surely linear mapping from ${\cal S}_\R={\cal S}(\R^d,\R^N)$ to the space of real valued measurable functions (random variables) on some probability space that fulfills the following properties:
\begin{enumerate}
\item (Temperedness) $f_n\to f$ in ${\cal S}_\R$ $\Rightarrow$ $\varphi(f_n)\stackrel{\cal L}{\to}\varphi(f)$;
\item (Covariance) $\varphi(f)\stackrel{\cal L}{=}\varphi(f_g)$ $\forall$ $f\in S_\R$, $g=\{\Lambda,a\}$, $\Lambda\in SO(d)$, $a\in \R^d$, $f_g(x)=\tau(\Lambda)f(\Lambda^{-1}(x-a))$ for some continuous representation $\tau:SO(d)\to GL(N)$.
\item  (Mixing) $\lim_{t\to\infty}\E[AB_{ta}]=\E[A]\E[B]$ for all square integrable random variables $A=A(\varphi), B=B(\varphi)$ and $B_{ta}=B(\varphi_{ta})$, $\varphi_{ta}(f)=\varphi(f_{ta})$ $\forall f\in S_\R$, $a\in\R^d\setminus \{0\}$.
\end{enumerate}
The mixing condition in the Euclidean space-time plays the same r\^ole as the cluster property in the generalized Wightman axioms.

In particular, we consider such random fields $\varphi$ obtained as solutions of the stochastic partial differential equation $D\varphi=\eta$. In this equation, $\eta$ is a noise field, i.e.
$\eta$ is $\tau$-covariant for some representation, $\eta(f)$ has infinitely divisible probability law and $\eta(f),\eta(h)$ are independent $\forall f,h\in{\cal S}_\R$ with $\mbox{supp}f\cap\mbox{supp}h=\emptyset$. $D$ is a $\tau$-covariant (i.e. $\tau(\Lambda) D \tau(\Lambda)^{-1}=D$ $\forall \Lambda\in SO(d)$) partial 
differential operator with constant coefficients (also pseudo differential operators $D$ could be considered). From the classification of infinitely divisible probability laws it is known that $\eta$ essentially consists of Gaussian white noise and Poisson fields and derivatives thereof. Such a Gauss-Poisson noise field by the Bochner--Minlos theorem is characterized by its Fourier transform. Direct relations with QFT arise if one chooses
\begin{equation}
\label{4.eqa}
\E[e^{i\eta(f)}]=\exp\left\{\int_{\R^d}\psi(f)-f\cdot \bar\sigma^2 p(-\Delta)f\,dx\right\},~~f\in{\cal S}_\R
\end{equation}
where $\psi:\R^N\to\C$ is a L\'evy function, 
\begin{equation}
\label{5eqa}
\psi(t)=ia\cdot t-{t\cdot\sigma^2 t\over 2}+z\int_{\R^N\setminus \{0\}}(e^{it\cdot s}-1)\, dr(s)\, ,~~t\in\R^N.
\end{equation}  
Here $\cdot$ is a $\tau$-invariant scalar product on $\R^N$, $\sigma$ a positive semidefinite $\tau$-invariant $N\times N$ matrix, $z\geq 0$ a real number and $r$ is a $\tau$-invariant probability 
measure on $\R^n\setminus\{0\}$ with all moments. $\bar \sigma^2_{\alpha,\beta}=(\partial^2\psi(t)/\partial t_\alpha\partial t_\beta)|_{t=0}$. $p:[0,\infty)\to[0,\infty)$ is a polynomial depending on $D$. If $\hat D^{-1}$, the Fourier transformed inverse of $D$, exists, it can be represented by 
\begin{equation}
\label{6eqa}
\hat D^{-1}(k)={Q_E(k)\over\prod_{l=1}^P(|k|^2+m_l^2)^{\nu_l}}
\end{equation} 
Here $Q_E(k)$ is a complex $N\times N$ matrix with polynomial entries being $\tau$-covariant, $\tau(\Lambda)Q_E(\Lambda^{-1}k)\tau(\Lambda)^{-1}=Q_E(k)$ $\forall \Lambda \in SO(d)$, $k\in\R^d$. $\nu_l\in\N$ and $m_l\in\C\setminus(-\infty,0)$ are parameters with the interpretation of the mass spectrum $(m_1,\ldots, m_P)$ and $(\nu_1,\ldots,\nu_P)$ the dipole degrees of the related masses. 
We restrict ourselves to the case of positive mass spectrum where $m_l>0$ and in this case
\begin{equation}
\label{7eqa}
p(t)=p(t,D)={\prod_{l=1}^P(t+m_l^2)^{\nu_l}\over\prod_{l=1}^P m_l^{2\nu_l}}\,, ~~t>0.
\end{equation}
One can show that $\varphi$ obtained as the unique solution of the SPDE $D\varphi=\eta$ is an Euclidean covariant, mixing random field. The Schwinger
functions (moments) of $\varphi$ are given by
\begin{equation}
\label{8}
S_{n}(f_1\otimes\cdots\otimes f_n)= E\left[\varphi(f_1)\cdots \varphi(f_n)\right],~f_1,\ldots f_n\in{\cal S}_\R.
\end{equation}
One then gets that
the Schwinger functions can be calculated explicitly. They are determined by the truncated Schwinger functions, cf. (\ref{4eqa}), as follows: 
For $n=2$ 
\begin{equation}
\label{10eqa}
S_{2,\alpha_1,\alpha_2}^T(x_1,x_2)={Q^E_{2,\alpha_1,\alpha_2}(-i\underline{\nabla}_2)\over \prod_{l=1}^Nm_l^{2\nu_l}}\left[\prod_{l=1}^N(-\Delta+m_l^2)^{-\nu_l}\right] (x_1-x_2)
\end{equation}
and for $n\geq 3$
\begin{eqnarray}
\label{11eqa}
S^T_{n,\alpha_1\cdots\alpha_n}(x_1,\ldots,x_n) &=& Q^E_{n,\alpha_1\cdots\alpha_n}(-i\underline{\nabla}_n)\nonumber\\
&\times&\int_{\R^d}\prod_{j=1}^n\left[\prod_{l=1}^N(-\Delta+m_l^2)^{-\nu_l}\right](x_j-x) \, dx
\end{eqnarray}
where
\begin{equation}
\label{12eqa}
Q^E_{n,\alpha_1\cdots\alpha_n}(-i\underline{\nabla}_n)=
C^{\beta_1\cdots\beta_n}\prod_{l=1}^nQ_{E,\beta_l,\alpha_l}(-i{\partial\over\partial x_l})
\end{equation}
with
\begin{equation}
\label{13eqa}
C_{\beta_1\cdots\beta_n}=(-i)^n\left.{\partial^n\psi(t)\over \partial t_{\beta_1}\cdots \partial t_{\beta_n}}\right|_{t=0}
\end{equation}
and the Einstein convention of summation and uppering/lowering of indices on 
$\R^N$ w.r.t. the invariant inner product $\cdot$ is applied. The Schwinger functions fulfill the requirements of $\tau$-covariance, symmetry, clustering and Hermiticity from the Osterwalder-Schrader
axioms of Euclidean QFT.

While there is no known general reason, why a relativistic QFT should exist for the given set of Schwinger functions, one can take advantage of the explicit formulae (\ref{10eqa})--(\ref{13eqa}) in order to
calculate the analytic continuation from Euclidean to relativistic times explicitly. 

It simplifies the considerations to exclude dipole fields, i.e. one assumes that $\nu_l=1$ for $l=1,\ldots,n$. In physical terms, the no-dipole condition guarantees that the asymptotic fields
in Minkowski space-time fulfill the Klein-Gordon equation and thus generate particles in the usual sense if applied to the vacuum. If this condition is not imposed, asymptotic fields might only fulfill a dipole equation $(\Box+m^2)^2\phi^{\rm in/out}=0$ or a related hyperbolic equation of even higher order and the
particle states generated by application of such fields to the vacuum require a gauge fixing (constraints) in order to obtain a physical interpretation. Given the no-dipole condition, one obtains by expansion into partial fractions
\begin{equation}
\label{14eqa}
{1\over
\prod_{l=1}^P(|k|^2+m_l^2)}=\sum_{l=1}^N
{b_{l}\over (|k|^2+m_l^2)}
\end{equation}
with $b_{l}\in (0,\infty)$ uniquely determined and $b_{l}\not =0$. For the truncated Schwinger functions this implies ($n\geq 3$)
\begin{eqnarray}
\label{15eqa}
S_{n,\alpha_1\cdots\alpha_n}^T(x_1,\ldots,x_n)&=&  Q_{n,\alpha_1
\cdots\alpha_n}^E(-i\underline{\nabla}_n)\sum_{l_1,\ldots,l_n=1}^P
\nonumber\\
&\times& \prod_{r=1}^nb_{l_r}\int_{\R^d}\prod _{j=1}^n(-\Delta+m_{l_j}^2)^{-1}(x-x_j)~dx.
\end{eqnarray}
At this point, a lengthy calculation yields a representation of the functions $\int_{\R^d}\prod _{j=1}^n(-\Delta+m_{j}^2)^{-1}(x-x_j)~dx$ as the Fourier--Laplace transform of a 
distribution $\hat W_{n,m_1,\ldots,m_n}^T$ that fulfills the spectral condition. This is equivalent to the statement that the analytic continuation of such functions to relativistic times yields $W_{n,m_1,\ldots,m_n}^T$, where the latter
distribution is the inverse Fourier transform of $\hat W_{n,m_1,\ldots,m_n}^T$. This distribution up to a constant that can be integrated into $Q^E$ is given by
\begin{equation}
\label{16eqa}
\left\{\sum_{j=1}^n\prod_{l=1}^{j-1}\delta^-_{m_l}(k_l){(-1)\over k^2-m_j^2}\prod_{l=j+1}^n\delta^+_{m_l}(k_l)\right\}\delta(\sum_{l=1}^nk_l)
\end{equation}
Here $\delta_m^{\pm}(k)=\theta (\pm k^0)\delta(k^2-m^2)$, where $\theta$ is the Heaviside step function and $k^2={k^0}^2-|\vec
k|^2$. On the other hand the partial differential operator $Q^E_n$ can be analytically continued in momentum space
\begin{equation}
\label{17eqa}
Q^M_{n}((k_1^0,\vec k_1),\ldots,
 (k_n^0,\vec k_n))=Q^E_{n}((ik_1^0,\vec k_1),\ldots,(i k_n^0,
\vec k_n)),
\end{equation}
$k_1,\ldots,k_n\in\R^d$. With the definition 
\begin{equation}
\label{18eqa}
\hat W_{2,\alpha_1\alpha_2}^T(k_1,k_2)=(2\pi)^{(d+1)}{Q^M_{2,\alpha_1\alpha_2}(k_1,k_2)\over
\prod_{l=1}^Nm_l^2}\sum_{l=1}^Nb_l\,\delta_{m_l}^-(k_1)\delta(k_1+k_2)
\end{equation}
and
\begin{eqnarray}
\label{19eqa}
\hat W_{n,\alpha_1\cdots\alpha_n}^T(k_1,\ldots,k_n)&=&Q^M_{n,\alpha_1\cdots\alpha_n}( k_1,\ldots,
 k_n)\nonumber\\
 &\times&\sum_{l_1,\ldots,l_n=1}^N\prod_{j=1}^nb_{l_j} \hat W_{n,m_{l_1},\ldots,m_{l_n}}^T(k_1,\ldots,k_n),
\end{eqnarray}
the analytic continuation of Schwinger functions can be summarized as follows:
\begin{theorem}
\label{2theo}
The truncated Schwinger functions $S_n^T$ have a Fourier-Laplace representation with
$\hat W_n^T$ defined in Eqs. (\ref{18eqa}) and (\ref{19eqa}).  Equivalently, $S_n^T$ is the analytic continuation of
$W_n^T$ from purely real relativistic time to purely imaginary Euclidean time.
The truncated Wightman functions $W_n^T$ fulfill the requirements of
temperedness, relativistic covariance w.r.t. the representation of the orthochronous, proper Lorentz group
$\tilde \tau: {\rm L}^\uparrow_+(d)\to {\rm Gl}(L)$, locality, spectral property and
cluster property.  Here $\tilde \tau$ is  obtained by analytic
 continuation of $\tau$
to a representation of the proper complex Lorentz group over $\, \C^d$
 (which contains $SO(d)$
as a real submanifold) and restriction of this representation to the
real orthochronous proper Lorentz group.
\end{theorem}
Making again use of the explicit formula in Theorem \ref{2theo}, the condition of Theorem \ref{1theo} can be verified. This proves the existence of IMQFT models associated to the class of random fields under discussion.
\begin{theorem}
\label{4theo}
The Wightman functions defined in Theorem \ref{2theo} fulfill the HSSC. In particular, there exists a QFT with indefinite metric s.t.
the Wightman functions are given as the vacuum expectation values of that IMQFT.
\end{theorem}
Theories as described in Theorem \ref{2theo} obviously have trivial scattering behavior if the noise field $\eta$ is Gaussian, i.e. if in (\ref{6eqa}) $z=0$. 
In the case when there is also a Poisson component in $\eta$, i.e. $z>0$, higher order tuncated Wightman functions do not vanish and such theories have non-trivial scattering. 

Before the scattering of the models can be discussed, some words need to be said about scattering in IMQFT in general. The scattering theory in axiomatic QFT, Haag-Ruelle theory, relies on positivity. In fact, one can show that
in the class of models under discussion the LSZ asymptotic condition is violated if dipole degrees of freedom are admitted. In that case more complicated asymptotic conditions have to be used. In any case,
Haag--Ruelle theory cannot be adapted to IMQFT. 

Nevertheless, asymptotic fields and states can be constructed in IMQFT if one imposes a no-dipole condition in a mathematically precise way. 
Then the LSZ asymptotic condition leads to the construction of mixed VEVs of asymptotic in- and out-fields with local fields. The collection of such VEVs is called the form factor functional. After constructing this collection of mixed VEVs,
one can try to check the HSSC for this functional and the obtains a Krein space representation for the algebra generated by in- local and out-fields.

Following this line, asymptotic in- and out- particle states can be constructed for the given mass spectrum $(m_1,\ldots,m_P)$. If $a^{{\rm in/out}\dagger}_{\alpha,l}(k)$, $l=1,\ldots,P$ denotes the creation operator for an incoming/outgoing particle with mass $m_l$, spin component $\alpha$ and energy-momentum $k$, the following scattering
can be derived for $r$ incoming particles with masses $m_{l_1},\ldots,m_{l_r}$ and $n-r$ outgoing particles with masses $m_{l_{r+1}},\ldots,m_{l_n}$
\begin{eqnarray}
\label{20eqa}
&&\left\langle a^{{\rm in}\dagger}_{\alpha_1,l_1}(k_1)\cdots a^{{\rm in}\dagger}_{\alpha_r,l_r}(k_r)\Omega,a^{{\rm out}\dagger}_{\alpha_{r+1},l_{r+1}}(k_{r+1})\cdots a^{{\rm out}\dagger}_{\alpha_n,l_n}(k_n)\Omega\right\rangle^T\nonumber\\
&=& -(2\pi) i Q^M_{\alpha_1,\ldots\alpha_n}(-k_1,\ldots,-k_r,k_{r+1},\ldots,k_n)\prod_{j=1}^n\delta_{m_{l_j}}^+(k_j)\,\delta (K^{\rm in}-K^{\rm out}).\nonumber \\
\end{eqnarray} 
$K^{\rm in/out}$ stand for the total energy-momentum of in- and out-particles, i.e. $K^{\rm in}=\sum_{j=1}^rk_j$ and $K^{\rm out}=\sum_{j=r+1}^nk_j$. 

Two immediate consequences can be drawn from (\ref{20eqa}). Firstly, choosing a model with non-vanishing Poisson part such that $C_{\beta_1\beta_2\beta_3}\not=0$ and a differential operator $D$ containing in its mass spectrum the masses $m$ and $\mu$ with $m>2\mu$ one gets a non-vanishing
scattering amplitude for the process
\begin{equation}
\label{21eqa}
\begin{array}{c}
\begin{picture}(100,50)(0,0)
\put (50,25){\circle{20}} \put(10,25){\line(1,0){30}}
\put(25,25){\vector(1,0){1}} \put(60,30){\vector(1,1){20}}
\put(60,20){\vector(1,-1){20}} \put(-2,22){$m$} \put(83,48){$\mu$}
\put(83,-1){$\mu$}
\end{picture}
\end{array}
\end{equation}
even though in- and out-particle states consist of particles with well-defined sharp masses. Thus, for the incoming particle the energy uncertainty which for a particle at rest is proportional to the mass uncertainty vanishes but still the particle undergoes a non-trivial decay and must have a finite decay time. This appears to be a contradiction
to the energy-time uncertainty relation, which therefore seems to have a unclear status in IMQFT, (i.e. in QFT including gauge fields). The origin of this inequality, which of course is experimentally very well tested, apparently has to be located in the constraints of the theory and not in the unconstrained IMQFT. 

Secondly, one can replace somewhat artificially the polynomials $Q^M_n$ in (\ref{17eqa}) by any other symmetric and relativistically covariant polynomial. If the sequence of the "new" $Q_n^M$ is of uniformly bounded degree in any of the arguments $k_1,\ldots k_n$, the re-defined Wightman functions in (\ref{17eqa}) still
fulfill the requirements of Theorem \ref{1theo} and thus define a new relativistic, local IMQFT. The scattering amplitudes of such a theory are again well-defined and given by (\ref{20eqa}). E.g. in the case of only one scalar particle with mass $m$ one can show that arbitrary Lorentz invariant scattering behavior of Bosonic particles
can be reproduced by such theories for energies below an arbitrary maximal energy up to arbitrary precision. This kind of interpolation theorem shows that the outcome of an arbitrary scattering experiment can be reproduced within the formalism of (unconstrained) IMQFT as long as it is in agreement with the general requirements of Poincar\'e
invariance and statistics.

\end{document}